\documentclass[fp,twocolumn]{jpsj3}

\usepackage{txfonts}

\usepackage{graphicx}
\usepackage{dcolumn}
\usepackage{bm}
\usepackage{color}
\usepackage{mathrsfs}

\title{ 
Frequency Dependence of Phonon-Induced Current Noise in Armchair Carbon Nanotube}
\author{Raimu Akimoto$^{1,*}$, Aina Sumiyoshi$^{1,*}$ and Takahiro Yamamoto$^{1,2}$}

\inst{$^1$ Department of Physics, Tokyo University of Science, 1-3 Kagurazaka, Shinjuku, Tokyo 162-8601, Japan \\
$^2$ RIST, Tokyo University of Science, 1-3 Kagurazaka Shinjuku, Tokyo 162-8601, Japan} %\\

\abst{
We theoretically investigate the frequency dependence of phonon-induced current noise in armchair carbon nanotubes at room temperature.
Our results reveal the emergence of multiple resonance peaks in the high-frequency regime, which cannot be accounted for by the Lorentzian 
lineshape expected from a Markovian process. The electron--phonon scattering processes responsible for most of these peaks are identified 
based on energy and momentum conservation laws and conventional selection rules. However, certain peaks cannot be fully explained within 
the framework of harmonic phonon scattering, suggesting the involvement of nontrivial interactions between electrons and anharmonic phonons.}

\begin{document}
\maketitle

\section{Introduction~\label{sec:1}}    
As emphasized by the title of Landauer's seminal article~\cite{rf:landauer}, {\it The noise is the signal}, electric current noise is a fundamental 
physical quantity that provides crucial insights into charge transport and scattering mechanisms in materials, and is therefore one of the central 
topics in mesoscopic physics~\cite{rf:Blanter,rf:kobayashi}. In particular, noise arising from phonons encodes detailed information about 
electron-phonon interactions and energy relaxation processes, shedding light on intrinsic material properties~\cite{rf:haupt,rf:chen}. 
Thus, the study of phonon-induced current noise holds significant importance both for deepening our understanding of fundamental physical 
phenomena and for advancing electronics.

In macroscopic bulk systems, current noise is often negligible; however, in small systems such as low-dimensional and nanoscale materials, 
noise effects become essential for understanding charge transport characteristics. Carbon nanotubes (CNTs), which are experimentally realizable 
one-dimensional (1D) nanomaterials~\cite{rf:iijima}, provide an ideal platform for investigating current noise in the nanoscale regime~\cite{rf:roche,rf:onac,rf:wu}. 
Recently, we conducted a theoretical investigation focusing on phonon-induced current noise in CNTs~\cite{rf:sumiyoshi}, with particular attention 
to the size dependence of the \textit{zero-frequency} noise power, namely, the power spectral density (PSD), $S(\omega = 0)$. We found that 
$S(0)$ exhibits a nontrivial dependence on the CNT length $L$. In the ballistic regime ($L \ll L_0$, where $L_0$ is the mean free path), $S(0)$ 
increases linearly with $L$; it reaches a maximum around $L \sim L_0$, and in the diffusive transport regime ($L \gg L_0$), it decreases following 
a power-law scaling, $S(0) \propto L^{-\alpha}$, with $\alpha > 0$. This characteristic length dependence represents a universal current-noise 
feature of 1D quantum transport governed by electron--phonon scattering, and is expected to offer new insights into nonequilibrium noise 
phenomena at the nanoscale.

Previous studies of the zero-frequency PSD, $S(0)$, have advanced our understanding of current noise~\cite{rf:Blanter,rf:kobayashi}. 
However, to fully understand its physical nature, it is essential to go beyond $S(0)$ and investigate the {\it frequency-dependent} PSD, 
$S(\omega)$. Indeed, $S(\omega)$ contains rich physical information that is inaccessible from $S(0)$ alone; for example, it has the potential 
to capture dynamical processes such as inter-valley scattering mediated by short-wavelength phonons in CNTs. Therefore, understanding 
the behavior of $S(\omega)$ is of fundamental interest in mesoscopic physics, particularly in the context of nonequilibrium processes and 
scattering mechanisms in nanomaterials. To the best of the authors' knowledge, no previous study has achieved a high-accuracy, 
atomistic-level analysis of the $\omega$-dependent current noise in realistic nanoscale materials such as CNTs.

In this study, we perform large-scale simulations of the $\omega$-dependent current noise $S(\omega)$ in metallic CNTs at room temperature, 
using our recently developed atomistic quantum transport method, the OpenTDSE+MD approach~\cite{rf:sumiyoshi,rf:ishizeki2017,rf:ishizeki2018,rf:ishizeki2020}. 
The simulations reveal multiple resonance peaks in the high-$\omega$ regime of $S(\omega)$, which cannot be explained by the Lorentzian lineshape of 
$S(\omega)$ predicted under the assumption that electron transport follows a simple Markovian process. In this paper, we explain the scattering 
mechanisms responsible for the emergence of these resonance peaks.

The paper is organized as follows. In Sec.~2, we describe the computational model and methods used in the present study. In Sec.~3, 
we present the simulation results and provide physical interpretations. The final section (Sec.~4) is devoted to the summary and outlook.

\section{Model and Methods}

\subsection{Basics of current noise}
Let us consider current-carrying steady states including noise. The magnitude of the current noise is characterized by 
the PSD~\cite{rf:Blanter,rf:kobayashi}, which is defined as
\begin{eqnarray}
S(\omega) = \lim_{T \to \infty} \frac{2}{T} \langle |\Delta J(\omega)|^2 \rangle.
\label{eq:power_spectrum01}
\end{eqnarray}
Here, $\langle \cdots \rangle$ denotes the ensemble average, taken over repeated experimental or computational realizations of 
the system, each performed under identical conditions, and $\Delta J(\omega)$ is the Fourier transform of the current fluctuation 
$\Delta J(t) = J(t) - \overline{J}$, given by
\begin{equation}
\Delta J(\omega) = \int_{-T/2}^{T/2} dt\, \Delta J(t) e^{i \omega t},
\label{eq:fourier}
\end{equation}
where $\overline{J}$ denotes the average current over the measurement interval $-T/2 \leq t \leq T/2$.
By substituting Eq.~(\ref{eq:fourier}) into Eq.~(\ref{eq:power_spectrum01}), 
$S(\omega)$ can be expressed as
\begin{eqnarray}
S(\omega) = 2 \int_{-\infty}^{\infty} d\tau\, C(\tau) e^{i \omega \tau}, \qquad \tau \equiv t - t',
\label{eq:power_spectrum03}
\end{eqnarray}
where $C(\tau)$ is the autocorrelation function of the current fluctuation, defined as
\begin{eqnarray}
C(\tau) = \lim_{T \to \infty} \frac{1}{T} \int_{-T/2}^{T/2} dt\, \langle \Delta J(t + \tau)\, \Delta J(t) \rangle,
\label{eq:auto_correlation}
\end{eqnarray}
in accordance with the Wiener--Khinchin theorem~\cite{rf:kampen}.

For simplicity, we now consider a situation in which atomic vibrations are completely random and the electronic transport process  
can be described by the simplest Markov process, namely a first-order autoregressive model (AR(1)). 
Within the AR(1) model, it is well known that the autocorrelation function 
$C(\tau)$ decays exponentially with the elapsed time $\tau$, and is given by
\begin{eqnarray}
C(\tau) = C(0) e^{-\tau / \tau_{\rm c}},
\label{eq:Ctau}
\end{eqnarray}
where $C(0) \equiv \langle \Delta J(0)^2 \rangle$ denotes the zero-time variance of the current fluctuation,  
and $\tau_{\rm c}$ is the characteristic decay time.
Substituting Eq.~(\ref{eq:Ctau}) into Eq.~(\ref{eq:power_spectrum03}) yields the Lorentzian form of the PSD:
\begin{eqnarray}
S(\omega) = \frac{S(0)}{1 + (\tau_{\rm c} \omega)^2}\equiv S_{\rm L}(\omega),
\label{eq:power_spectrum04}
\end{eqnarray}
where $S(0) \equiv 4 C(0) \tau_{\rm c}$. As seen in Eq.~(\ref{eq:power_spectrum04}), it behaves as white noise ($S(\omega) = \text{const.}$) 
in the low-$\omega$ regime ($\omega \ll \omega_{\rm c}$) and decays as $1/\omega^2$ in the high-$\omega$ regime ($\omega \gg \omega_{\rm c}$). 
The cutoff frequency is defined as $\omega_{\rm c} \equiv 1/\tau_{\rm c}$, where the white noise intensity is reduced by half. 

In Sec.~3, we use Eq.~(\ref{eq:power_spectrum04}) to analyze the simulation results for current noise in CNTs. 

\subsection{Simulation setup}
As shown in Eqs.~(\ref{eq:power_spectrum01}) and (\ref{eq:fourier}), the PSD $S(\omega)$ can be evaluated once the time-dependent 
current $J(t)$ is obtained. To compute $J(t)$ for a CNT under electron--phonon scattering, we consider a two-terminal configuration 
in which a central CNT of finite length $L$ is connected to CNT leads of semi-infinite length, each having the same chirality as 
the central region. Scattering events are included only within the central CNT, 
while the left and right leads are assumed to be defect free and non interacting. In this setup, electrons injected from one lead propagate through 
the central CNT while undergoing scattering with phonons, and are eventually transmitted into the opposite lead. In the absence of electron--phonon scattering, current noise vanishes. By calculation the rate of transmitted electrons as a function of time, the time-dependent current $J(t)$ can be evaluated.

A powerful framework for simulating quantum transport in such a setup is the OpenTDSE+MD method, which has been recently developed 
by our group~\cite{rf:sumiyoshi,rf:ishizeki2017,rf:ishizeki2018,rf:ishizeki2020}. As the full details of the method are given in Refs.~[\citen{rf:sumiyoshi,rf:ishizeki2017,rf:ishizeki2018,rf:ishizeki2020}], we briefly summarize its key features below and describe 
the specific simulation conditions used in the present study. In the OpenTDSE+MD method, the nuclear dynamics in the central CNT 
at finite temperature are simulated using classical molecular dynamics (MD) under the NTV ensemble~\cite{rf:NTV}
with velocity scaling method~\cite{rf:velocity01,rf:velocity02}. 
The interatomic forces between carbon atoms in the CNT are modeled using the Tersoff potential~\cite{rf:lindsay}, which is known to provide 
a reliable description of carbon--carbon interactions. The effect of nuclear motion on the electronic structure is 
incorporated into the time-dependent hopping integrals $\gamma_{ij}(t)$ of the tight-binding Hamiltonian, following the Harrison 
rule~\cite{rf:harrison}:
\begin{equation}
\gamma_{ij}(t) = \gamma_0 \frac{|{\bm R}_{0,i} - {\bm R}_{0,j}|^2}{|{\bm R}_i(t) - {\bm R}_j(t)|^2},
\label{eq:hopping}
\end{equation}
where $\gamma_0 = -2.7$~eV is the equilibrium hopping integral, ${\bm R}_{0,i}$ is the equilibrium position of the $i$th carbon atom, 
and ${\bm R}_i(t)$ is its position at time $t$, obtained from the MD simulation. By solving the time-dependent Schr{\" o}dinger equation 
(TDSE) under the two-terminal configuration using the hopping integrals defined in Eq.~(\ref{eq:hopping}), we obtain the time-dependent 
wavefunction, which is subsequently used to calculate the instantaneous current through the central CNT.

In this study, we focus on armchair single-walled CNTs (SWCNTs) with various diameters and a fixed length $L=L_0$, where the noise 
reaches its maximum~\cite{rf:sumiyoshi}. The Fermi energy is set at the charge neutral point, which is taken as the energy of the incident electron.

\section{Simulation Results and Discussion}

\subsection{$S(\omega)$ of phonon-induced current noise}
\begin{figure}[t]
\begin{center}
\includegraphics[width=80mm]{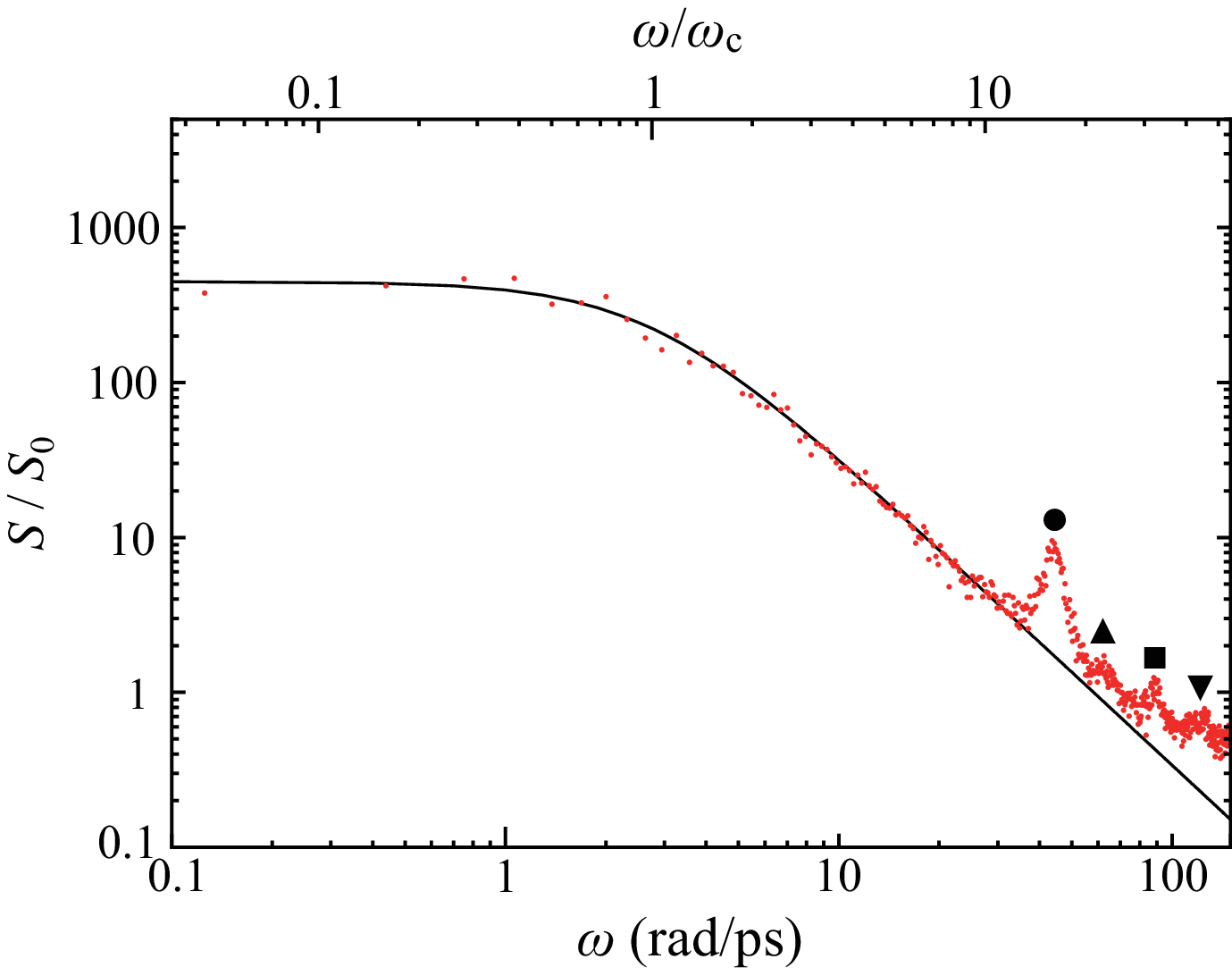}
\caption{(Color online)
Frequency dependence of the current noise PSD, $S(\omega)$, for a (7,7) SWCNT with length $L = 430$~nm at 300~K.  
Here, $S_0 \equiv J_0^2 \Delta t$ is a characteristic PSD, where $J_0 \equiv 2e^2V/h$ and $\Delta t = 0.1$~fs is the time step. 
The solid curve represents a Lorentzian-shaped PSD described by Eq.~(\ref{eq:power_spectrum04}). In the high-$\omega$ region, 
the four prominent deviation points ($\bullet$, $\blacktriangle$, $\blacksquare$, and $\blacktriangledown$) are observed.
}
\label{fig:01}
\end{center}
\end{figure}

Figure~\ref{fig:01} shows $S(\omega)/S_0$ for the (7,7) SWCNT at the room temperature (300K), whose length $L$ equals to 
the mean free path $L_0 = 430$~nm~\cite{rf:mfp}. 
Here, $S_0 \equiv J_0^2 \Delta t$ is a characteristic PSD, where $J_0 \equiv 2e^2V/h$ is the electric current through a single channel with 
perfect transmission ($e$ is the elementary charge and $V$ is the bias voltage between the two leads), and $\Delta t$ is the time step, 
set to $\Delta t = 0.1$~fs. The measurement interval for each trial was set to $T = 0.1$~ns, which is much longer than the mean free time 
$\tau_{\rm ph} (= L_0 / v_{\rm F}) = 0.3$~ps. The red dots in Fig.~\ref{fig:01} represent the ensemble average of the PSD over $N = 25$ trials:
\begin{equation}
\left\langle S(\omega) \right\rangle = \frac{1}{N} \sum_{k=1}^{N} S_k(\omega),
\label{eq:}
\end{equation}
where $k$ indexes each trial, and a sectional (bin-wise) average was performed over groups of 5 adjacent frequency points to smooth 
the spectra. The black curve represents a Lorentzian-shaped PSD described by Eq.~(\ref{eq:power_spectrum04}), which assumes 
a Markovian process with random scattering by phonons. 
In the low- and mid-$\omega$ regions, the Lorentzian-shaped PSD with $\tau_{\rm c}=0.36$~ps ($\omega_{\rm c}=2.78$~THz) 
agrees well with the OpenTDSE-MD results (the red dots). This suggests that the current noise in these frequency ranges originates 
from random scattering by phonons, specifically the twisting (TW) acoustic phonons~\cite{rf:suzuura, rf:MSD, rf:Jiang, rf:popov}.

The characteristic features of white noise, particularly $S(0)$, have been investigated in detail in our previous work~\cite{rf:sumiyoshi}. 
Therefore, in this study, we turn our attention to the high-$\omega$ regime ($\omega \gg \omega_{\rm c}$). In this high-$\omega$ regime, 
obvious deviation from the Lorentzian background (the black curve) are observed, arising from electron--phonon scattering. Among them, 
we focus on four prominent deviation points indicated by $\bullet$, $\blacktriangle$, $\blacksquare$, and $\blacktriangledown$ in Fig.~\ref{fig:01}. 
The phonon modes responsible for these four resonance peaks are discussed in the following section (\S 3.2).

\subsection{Origin of resonant peaks appearing in $S(\omega)$}

\begin{figure}[t]
\begin{center}
\includegraphics[width=80mm]{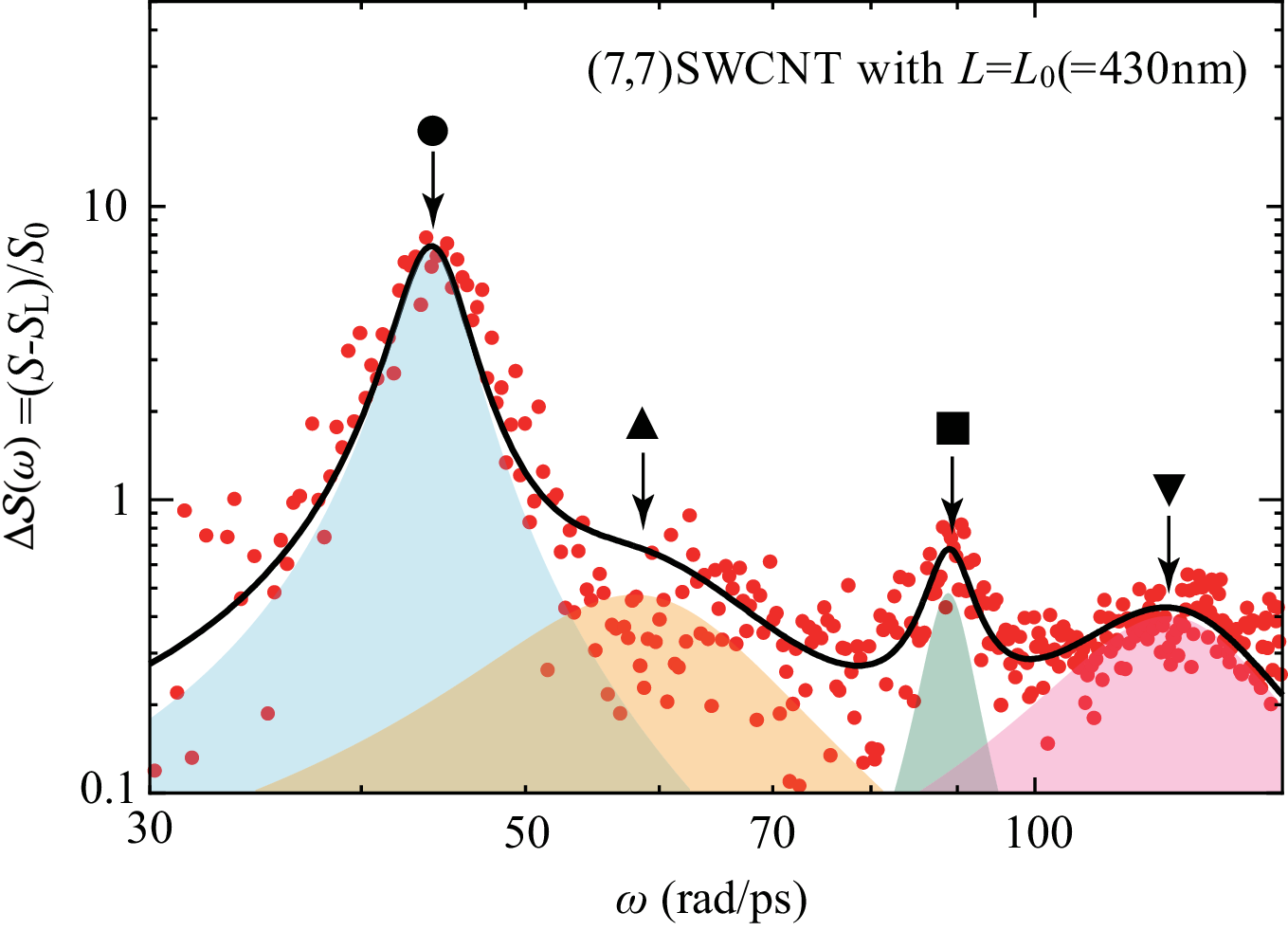}
\caption{(Color online)
Residual PSD $\Delta{\mathcal S}(\omega)=\{S(\omega) - S_{\rm L}(\omega)\}/S_0$ of the (7,7) SWCNT with $L=430$nm at 300K. }
\label{fig:02}
\end{center}
\end{figure}

Figure~\ref{fig:02} displays the residual PSD $\Delta{\mathcal S}(\omega) = \{S(\omega) - S_{\rm L}(\omega)\}/S_0$ in the high-$\omega$ region, 
obtained by subtracting the black curve shown in Fig.~\ref{fig:01} from the raw data. This residual PSD is well described by a sum of four individual 
Lorentzian functions:
\begin{equation}
\Delta{\mathcal S}(\omega) = \sum_{n=1}^4 A_n \frac{\gamma_n/2}{(\omega - \omega_n)^2 + (\gamma_n/2)^2},
\label{eq:Lorentzian}
\end{equation}
where $n = 1$, 2, 3, and 4 correspond to $\bullet$, $\blacktriangle$, $\blacksquare$, and $\blacktriangledown$.
$\omega_n$, $A_n$, and $\gamma_n$ represent the center frequency, amplitude, and linewidth of the $n$th Lorentzian peak, respectively. 
These parameters are obtained by fitting the data (red circles in Fig.~\ref{fig:02}) using Eq.~(\ref{eq:Lorentzian}), and are summarized in 
Table~\ref{tab:parameters}. In Fig.~\ref{fig:02}, the fitted Lorentzian components are depicted in translucent blue, yellow, green, and pink, respectively.
The coefficient of determination for the overall fitted curve (black curve in Fig.~\ref{fig:02}) is $R^2 = 0.91$, which is very close to unity, indicating that 
the model provides a good fit to the data. In the following, we attempt to identify the phonon modes corresponding to these four Lorentzian peaks and 
to elucidate the mechanisms responsible for their emergence.

\begin{table}[t]
 \caption{Parameters obtained by fitting the data (red circles in Fig.~\ref{fig:02}) using Eq.~(\ref{eq:Lorentzian}).
 }
\label{tab:parameters}
\begin{center}
\begin{tabular}{cccc}
\hline
 $n$ & $\omega_n$ (rad/ps) & $A_n$ & $\gamma_n$ (rad/ps) \\
\hline
    1 ($\bullet$)   & 44   & 16.0 &  4.5 \\
    2 ($\blacktriangle$) & 58   & 5.8 & 24.5  \\       
    3 ($\blacksquare$) & 89   & 1.6 & 6.5 \\
    4 ($\blacktriangledown$) & 120   & 8.0 & 40.0 \\      
\hline
\end{tabular}
\end{center}
\end{table}

\begin{figure}[t]
\begin{center}
\includegraphics[width=85mm]{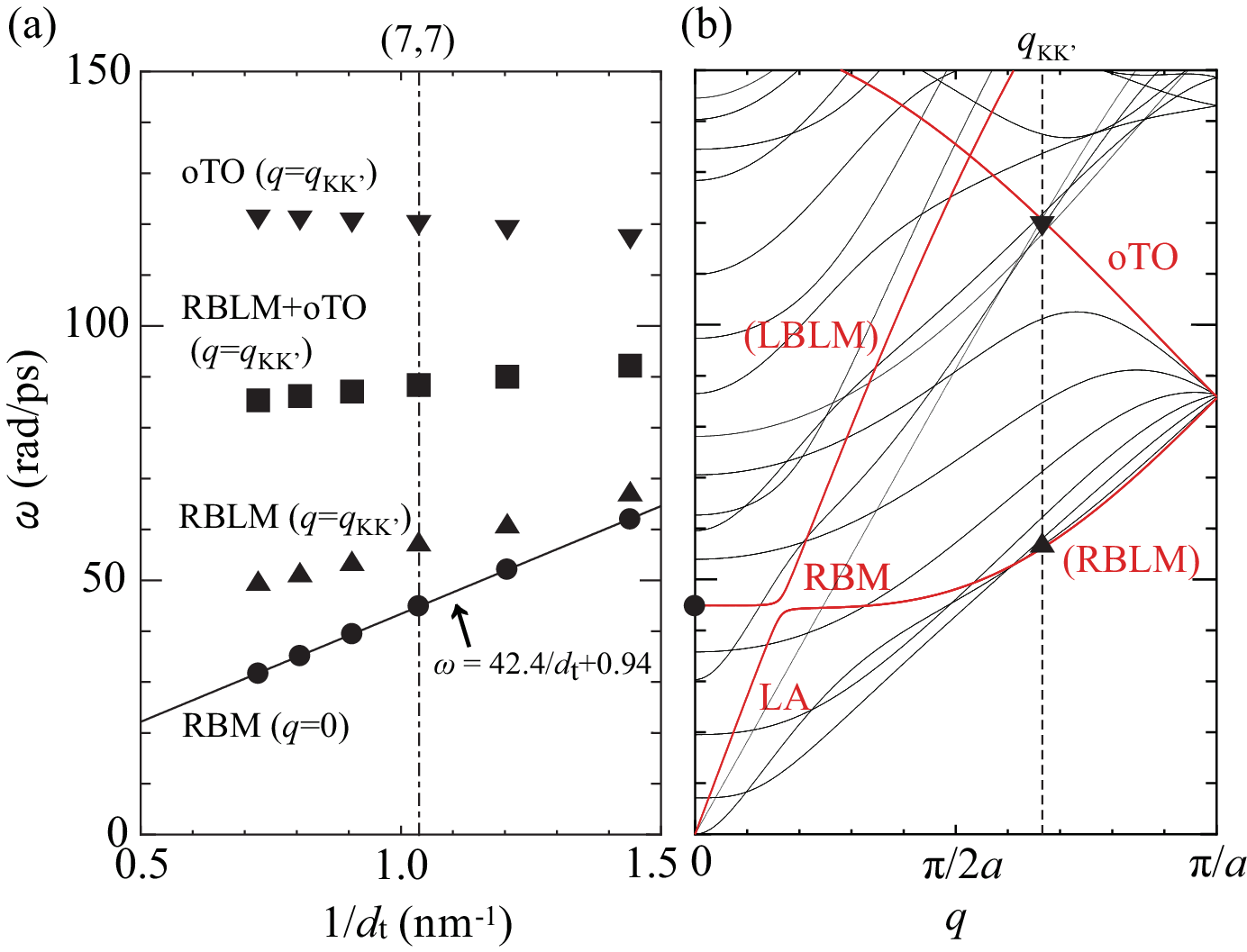}
\caption{(Color online)
(a) Dependence of the four resonance peaks ($\bullet$, $\blacktriangle$, $\blacksquare$, and $\blacktriangledown$) on the tube diameter 
$d_{\mathrm{t}}$ for $(n,n)$ SWCNTs with $n= 5\text{--}10$, assuming the tube length equals the mean free path. The corresponding diameters 
are 0.69, 0.83, 0.97, 1.10, 1.24, and 1.38~nm for $n = 5\text{--}10$, respectively.  (b) Phonon dispersion of the (7,7) SWCNT with unit cell length 
$a = 0.25$~nm. The vibrational characters of the longitudinal acoustic (LA) and RBM phonons are interchanged 
between the low- and high-$q$ regions after the anticrossing of two phonon branches, evolving into the RBLM and longitudinal breathing-like mode 
(LBLM) phonons, respectively.
}
\label{fig:03}
\end{center}
\end{figure}

\begin{figure}[t]
\begin{center}
\includegraphics[width=75mm]{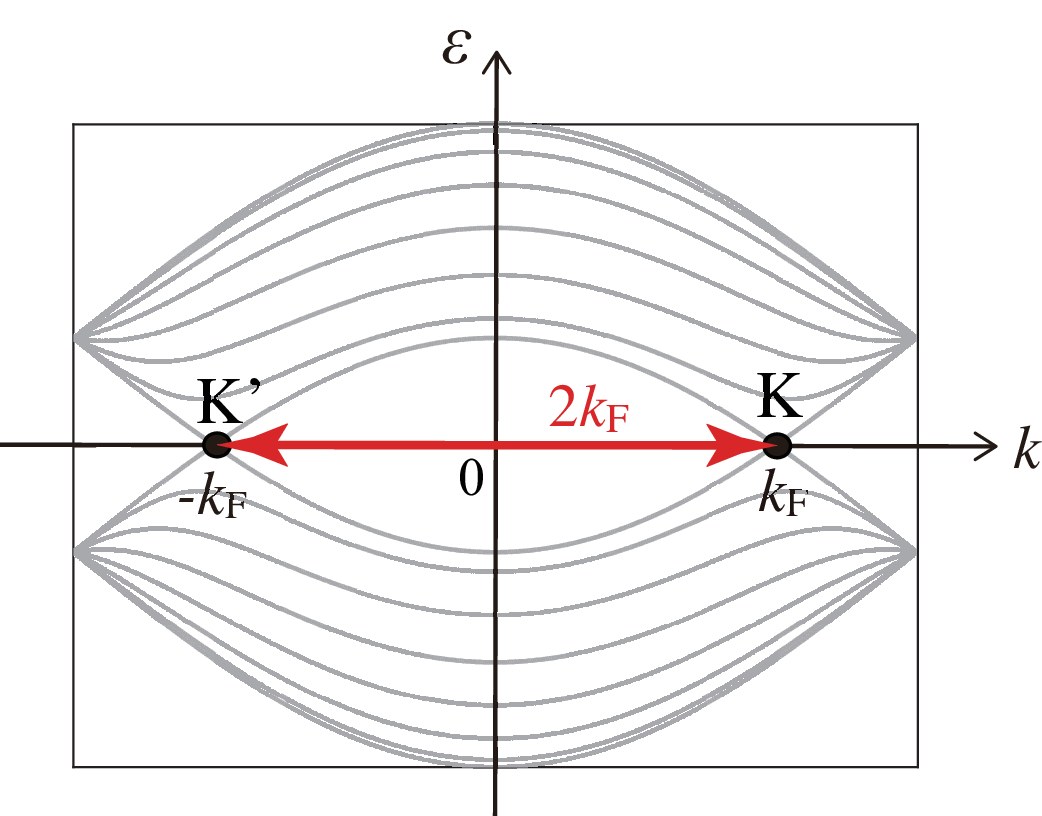}
\caption{(Color online)
Energy band near the Fermi energy ($\varepsilon_{\rm F} = 0$) for the (7,7) SWCNT.  The red double arrow indicates inter-valley scattering 
with momentum transfer $2k_{\rm F}$ between the two Fermi points $\pm k_{\rm F}$, corresponding to the K and K$'$ valleys.
}
\label{fig:04}
\end{center}
\end{figure}

Figure~\ref{fig:03}(a) displays the tube-diameter ($d_{\rm t}$) dependence of the four peak positions, obtained by performing similar calculations 
of $S(\omega)$ for various $(n,n)$ SWCNTs with $L=L_0$ at 300K~\cite{rf:mfp}.
First, we found that the $\bullet$ peak decreases inversely proportional to $d_{\rm t}$ and it can be fitted by $\omega = A/d_{\rm t} + B$, 
with $A = 42.4$~nm$\cdot$rad/ps and $B = 0.94$~rad/ps. This $d_{\rm t}$ dependence is consistent with the well-known diameter 
dependence of the radial breathing mode (RBM) phonon frequency at zero wavenumber ($q = 0$)~\cite{rf:mahan, rf:perebeinos, rf:jorio01}.  
Note that the small offset $B$ arises from the curvature effect of the C--C bonds in small-diameter SWCNTs.~\cite{rf:kurti, rf:jorio02}
The strong agreement suggests that the resonance peak ($\bullet$) originates from intra-K (or intra-K$'$) valley scattering of electrons mediated by 
an RBM phonon with $q = 0$. Here, we refer to the two charge neutral points as the K and K$'$ points, in analogy with the Dirac points in graphene, 
to which they correspond in the unrolled Brillouin zone~\cite{rf:ando_JPSJ} (See Fig.~\ref{fig:04}). This interpretation is also physically reasonable 
in light of the selection rules for electron--phonon scattering in $(n,n)$ SWCNTs~\cite{rf:MSD, rf:Jiang, rf:popov}.

Next, we consider the origin of the other resonance peaks from the viewpoint of the selection rules~\cite{rf:MSD, rf:Jiang, rf:popov}.
An electron in an $(n,n)$ SWCNT can couple to the radial breathing-like mode (RBLM) phonon and the
out-of-plane transverse optical (oTO) phonon with wavenumber $q = q_{\rm KK'}\equiv |2k_{\rm F}-G|$ as shown in Fig.~\ref{fig:03}(b).
Here, $G=2\pi/a$ is a reciprocal lattice vector and $k_{\rm F}=2\pi/3a$ is the Fermi wavevector of a charge-neutral $(n,n)$ SWCNT, and $2k_{\rm F}$ 
corresponds to the wavenumber connecting the K and K$'$ points (Fig.~\ref{fig:04}). Here, $a$(=0.25~nm) is the unit cell length for $(n,n)$ SWCNTs. 
Because these electron-phonon couplings cause the K-K$'$ inter-valley scattering, we can expect that 
$S(\omega)$ exhibits resonance peaks at frequencies $\omega = \omega_{{\rm RBLM}}$ and $\omega = \omega_{{\rm oTO}}$ with $q =q_{\rm KK'}$.
Indeed, as shown in Fig.~\ref{fig:02}, two peaks are clearly observed at $\omega = 58~\mathrm{rad/ps}$ ($\blacktriangle$) and $120~\mathrm{rad/ps}$ ($\blacktriangledown$), which are in close agreement with the expected mode frequencies $\omega_{\mathrm{RBLM}} = 56.4~\mathrm{rad/ps}$ and 
$\omega_{\mathrm{oTO}} = 120.4~\mathrm{rad/ps}$ of the (7,7) SWCNT.
Moreover, we calculated the $\blacktriangle$ and $\blacktriangledown$ peak positions for various $d_{\mathrm{t}}$, as shown in Fig.~\ref{fig:03}(a). These $d_{\mathrm{t}}$ dependences are found to be consistent with these of the RBLM and oTO phonon frequencies at 
$q=q_{\mathrm{KK'}}$. These results strongly suggest that the two resonance peaks ($\blacktriangle$ and $\blacktriangledown$) 
originate from K-K$'$ inter-valley electron scattering mediated by RBLM and oTO phonons with $q = q_{\mathrm{KK}'}$, respectively.

Finally, we discuss the remaining peak ($\blacksquare$) fitted by Lorentzian centered at 89~rad/ps in Fig.~\ref{fig:02}. Unlike the 
other three peaks ($\bullet$, $\blacktriangle$, and $\blacktriangledown$), this peak cannot be identified by any simple harmonic phonon scattering 
process that satisfies energy and momentum conservation as well as the selection rules. Therefore, we speculate that the appearance of this peak 
arises from anharmonic phonon effects.

In this scenario, a possible scattering mechanism involves a four-phonon process, in which two phonons (RBLM with $q = q_{\rm KK'}$ 
and $\omega_{\rm RBLM} = 58~\mathrm{rad/ps}$, and the oTO mode with $q = q_{\rm KK'}$ and $\omega_{\rm oTO} = 120~\mathrm{rad/ps}$) are 
annihilated, and two new phonons with $q = q_{\rm KK'}$ and $\omega = (\omega_{\rm oTO} + \omega_{\rm RBLM})/2$ are created.  In fact,
the $\blacksquare$ peak appears in the vicinity of $\omega = (\omega_{\rm oTO} + \omega_{\rm RBLM})/2$, supporting this scenario. From a 
symmetry standpoint, coupling between these two modes are allowed, as both phonons exhibit radial vibrational character. 
The newly generated phonons could contribute to inter-valley electron scattering.  
This interpretation is further supported by numerical observations: while $\omega_{\rm RBLM}$ and $\omega_{\rm oTO}$ exhibit positive and 
negative shifts, respectively, with increasing $d_{\rm t}$, the $\blacksquare$ peak remains near their average value, as shown 
in Fig.~\ref{fig:03}(a). These results suggest that the $\blacksquare$ peak originates from a four-phonon anharmonic process, highlighting 
the nontrivial role of phonon--phonon coupling beyond the linear scattering regime. A more sophisticated theoretical analysis would be 
desirable to fully validate this interpretation; however, such an investigation lies beyond the scope of the present work and is left for future studies.

\section{Summary and outlook}
We investigated the $\omega$ dependence of the PSD of phonon-induced current noise in $(n,n)$ SWCNTs, focusing especially on 
the (7,7) structure. In the low-$\omega$ regime, the PSD follows a Lorentzian form derived from a Markovian model. In the high-$\omega$ 
regime, four Lorentzian resonance peaks ($\bullet$, $\blacktriangle$, $\blacksquare$, and $\blacktriangledown$) appear, deviating from 
the Lorentzian lineshape. The $\bullet$ peak at 44~rad/ps is attributed to intra-valley scattering via the RBM at $q = 0$, whereas
the $\blacktriangle$ and $\blacktriangledown$ peaks at 58 and 120~rad/ps arise from inter-valley scattering mediated by 
RBLM and oTO phonons at $q = q_{\rm KK'}$, respectively. In contrast, the $\blacksquare$ peak at 89~rad/ps cannot be explained by harmonic 
processes and is likely due to an four-phonon anharmonic interaction involving RBLM and oTO phonon modes. Its frequency closely matches 
$(\omega_{\mathrm{RBLM}} + \omega_{\mathrm{oTO}})/2$ and shows consistent $d_{\rm t}$ scaling. These findings reveal that 
high-$\omega$ current noise in SWCNTs is shaped by both linear and nonlinear e-ph interactions, reflecting the complexity of carrier 
and phonon dynamics in low-dimensional systems.

Finally, we comment on the experimental feasibility of measuring the predicted effects. Conventional electronic measurement techniques 
currently lack the temporal resolution required to detect terahertz (THz) frequency noise. Recent advances in techniques such as noise 
correlation spectroscopy and single-shot spectroscopy have enabled the observation of fluctuations in the THz regime. These include reports on 
electric-field vacuum fluctuations~\cite{rf:ex01}, spin fluctuations~\cite{rf:ex02}, and incoherent fields emitted from resonant tunneling diode (RTD) 
devices. In addition, transient current responses on the picosecond timescale have been observed in on-chip devices based on high-mobility 
graphene~\cite{rf:ex03,rf:ex04}. These developments indicate a promising prospect for experimentally detecting current noise in the THz regime 
from carbon nanotubes (CNTs) under applied DC bias in the near future.

\section*{Acknowledgements}
We are grateful to Kazuhiro Yanagi and Satoshi Kusaba for helpful discussions regarding the experimental feasibility of 
the present theoretical predictions. This work was supported in part by the Japan Society for the Promotion of Science 
KAKENHI (grant no. 23H00259) and by Toshiba Electronic Devices \& Storage Corporation under an Academic Encouragement Grant.

\vspace{1.0cm}
\noindent * These authors contributed equally to this work.\\


\begin{thebibliography}{99}
\bibitem{rf:landauer} R. Landauer, Nature {\bf 392}, 658 (1998).
\bibitem{rf:Blanter} Y. M. Blanter and M. B{\" u}ttiker, Phys. Rep. {\bf 336}, 1 (2000).
\bibitem{rf:kobayashi} K. Kobayashi and M. Hashisaka, J. Phys. Soc. Jpn. {\bf 90}, 102001 (2021).
\bibitem{rf:haupt} F. Haupt, T. Novotn\'y and W. Belzig, Phys. Rev. B {\bf 82}, 165441 (2010).
\bibitem{rf:chen} Y.-C. Chen and M. Di Ventra, Phys. Rev. Lett. {\bf 95}, 166802 (2005).
\bibitem{rf:iijima} S. Iijima and T. Ichihashi, Nature, {\bf 363}, 603 (1993).
\bibitem{rf:roche} P.-E. Roche, M. Kociak, S. Gueron, A. Kasumov, B. Reulet and H. Bouchiat, Eur. Phys. J. B {\bf 28}, 217 (2002).
\bibitem{rf:onac} E. Onac, F. Balestro, B. Trauzettel, C. F. J. Lodewijk and L. P. Kouwenhoven, Phys. Rev. Lett. {\bf 96}, 026803 (2006).
\bibitem{rf:wu} F. Wu, P. Virtanen, S. Andresen, B. Placais and P. J. Hakonen, Appl. Phys. Lett. {\bf 97}, 262115 (2010).
\bibitem{rf:sumiyoshi} A. Sumiyoshi and T. Yamamoto, J. Phys. Soc. Jpn. (in press).
\bibitem{rf:ishizeki2017} K. Ishizeki, K. Sasaoka, S. Konabe, S. Souma and T. Yamamoto Phys. Rev. B {\bf 96}, 035428 (2017).
\bibitem{rf:ishizeki2018} K. Ishizeki, K. Sasaoka, S. Konabe, S. Souma and T. Yamamoto, Jpn. J. Appl. Phys. {\bf 57}, 065102 (2018).
\bibitem{rf:ishizeki2020} K. Ishizeki, K. Takashima, K. Sasaoka and T. Yamamoto, Jpn. J. Appl. Phys. {\bf 59}, 055001 (2020).
\bibitem{rf:kampen} N.G. Van Kampen., Stochastic Processes in Physics and Chemistry, 3rd Edition, North-Holland (2007).
\bibitem{rf:NTV} L. V. Woodcock, Chem. Phys. Lett. {\bf 10}, 257 (1971).
\bibitem{rf:velocity01} J. M. Haile and S. Gupta, J. Chem. Phys. {\bf 79}, 3067 (1983).
\bibitem{rf:velocity02} H. C. Anderson, J. Chem. Phys. {\bf 72}, 2384 (1980).
\bibitem{rf:lindsay} L. Lindsay and D. A. Broido: Phys. Rev. B {\bf 81}, 205441 (2010).
\bibitem{rf:harrison} W. A. Harrison: Electronic Structure and the Properties of Solids: The Physics of the Chemical Bond (Dover, New York, 1989).
\bibitem{rf:mfp} The calculated mean free paths $L_0$ for the $(n,n)$ SWCNTs with $n = 5$ to $10$ are found to be $L_0 = 306$, $368$, $430$, $488$, $550$, and $612$~nm, respectively. These results exhibit excellent agreement with previously reported theoretical predictions.~\cite{rf:ishizeki2017, rf:ishii, rf:suzuura}
\bibitem{rf:ishii} H. Ishii, N. Kobayashi, and K. Hirose, Phys. Rev. B {\bf 82}, 085435 (2010).
\bibitem{rf:suzuura} H. Suzuura and T. Ando, Phys. Rev. B 65, 235412 (2002).
\bibitem{rf:MSD} M. S. Dresselhaus, G. Dresselhaus, R. Saito, and A. Jorio, Phys. Rep. 409, 47 (2005).
\bibitem{rf:Jiang} J. Jiang, R. Saito, G. G. Samsonidze, S. G. Chou, A. Jorio, G. Dresselhaus, and M. S. Dresselhaus, Phys. Rev. B {\bf 72}, 235408, (2005).
\bibitem{rf:popov} V. N. Popov and P.Lambin, Phys. Rev. B {\bf 74}, 075415, (2006).
\bibitem{rf:mahan} G. D. Mahan, Phys. Rev. B {\bf 65}, 235402 (2002).
\bibitem{rf:perebeinos} V. Perebeinos and J. Tersoff, Phys. Rev. B {\bf 79}, 241409 (2009).
\bibitem{rf:jorio01} A. Jorio, R. Saito, J. H. Hafner, C. M. Lieber, M. Hunter, T. McClure, G. Dresselhaus, and M. S. Dresselhaus, Phys. Rev. Lett. {\bf 86}, 1118 (2001).
\bibitem{rf:kurti} J. K{\" u}rti, V. Z\'olyomi, M. Kertesz, and G. Y. Sun, New J. Phys. {\bf 5}, 125 (2003).
\bibitem{rf:jorio02} A. Jorio, C. Fantini, M. A. Pimenta, R. B. Capaz, Ge. G. Samsonidze, G. Dresselhause, M. S. Dresselhause, J. Jiang, N. Kobayashi, A. Gr{\" u}neis, and R. Saito, Phys. Rev. B {\bf 71}, 075401 (2005).
\bibitem{rf:ando_JPSJ} T. Ando, J. Phys. Soc. Jpn. {\bf 74}, 777 (2005).
\bibitem{rf:ex01} C. Riek, D. V. Seletskiy, A. S. Moskalenko, J. F. Schmidt, P. Krauspe, S. Eckart, S. Eggert, G. Burkard, and A. Leitenstorfer, Science {\bf 350}, 420 (2015).
\bibitem{rf:ex02} M. A. Weiss, A. Herbst, J. Schlegel, T. Dannegger, M. Evers, A. Donges, M. Nakajima, A. Leitenstorfer, S. T. B. Goennenwein, U. Nowak, and T. Kurihara, Nature Communication {\bf 14}, 7651 (2023).
\bibitem{rf:ex03} K. Yoshioka, T. Wakamura, M. Hashisaka, K. Watanabe, T. Taniguchi, and N. Kumada, Nature Photonics {\bf 16}, 718 (2022).
\bibitem{rf:ex04} K. Yoshioka, G. Bernard, T. Wakamura, M. Hashisaka, K. Sasaki, S. Sasaki, K. Watanabe, T. Taniguchi, and N. Kumada, Nature Electronics {\bf 7}, 537 (2024).
\end{thebibliography}
\end{document}